\documentclass[aip,apl,reprint,showpacs,10pt,citeautoscript]{revtex4-1}
\usepackage{subfigure}
\usepackage{graphicx}
\usepackage{dcolumn}
\usepackage{bm}
\usepackage{textcomp}
\usepackage{amsmath}
\usepackage{amssymb}
\usepackage{epstopdf}
\usepackage{natmove}
\usepackage{array}

\usepackage{soul}
\usepackage{color}


\begin{document}

\title{Defect identification based on first-principles calculations for deep level transient spectroscopy}

\author{Darshana Wickramaratne}
\affiliation{Materials Department, University of California Santa Barbara, California 93106-5050, USA}
\author{Cyrus E. Dreyer}
\affiliation{Department of Physics and Astronomy, Stony Brook University, Stony Brook, New York 11794-3800, USA}
\affiliation{Center for Computational Quantum Physics, Flatiron Institute, 162 5$^{th}$ Avenue, New York, New York 10010, USA}
\author{Bartomeu Monserrat}
\affiliation{TCM Group, Cavendish Laboratory, University of Cambridge, J. J. Thomson Avenue, Cambridge CB3 0HE, United Kingdom}
\author{Jimmy-Xuan Shen}
\affiliation{Department of Physics, University of California Santa Barbara, California 93106-9530, USA}
\author{John L. Lyons}
\affiliation{Center for Computational Materials Science, US Naval Research Laboratory, Washington, DC 20375, USA}
\author{Audrius Alkauskas}
\affiliation{Center for Physical Sciences and Technology (FTMC), Vilnius LT-10257, Lithuania}
\author{Chris G. Van de Walle}
\affiliation{Materials Department, University of California Santa Barbara, California 93106-5050, USA}

\begin{abstract}
Deep level transient spectroscopy (DLTS) is used extensively
to study defects in semiconductors.
We demonstrate that great care should be exercised in interpreting activation energies extracted from DLTS as ionization energies.
We show how first-principles calculations of thermodynamic transition levels, temperature effects of ionization energies, and nonradiative capture coefficients can be
used to accurately determine actual activation energies that can be directly compared with DLTS.
Our analysis is illustrated with hybrid functional calculations for two important defects in GaN that have similar
thermodynamic transition levels, and shows that
the activation energy extracted from DLTS includes a capture barrier that is temperature dependent, unique to each defect, and in some cases large in comparison to the ionization energy.
By calculating quantities that can be directly compared with experiment, first-principles calculations thus offer powerful leverage in identifying the microscopic origin of defects detected in DLTS.
\end{abstract}
\date{\today}

\maketitle

Point defects and impurities are present in all semiconductors. They
can act as recombination centers that lower the efficiency of
optoelectronic devices,
or as carrier traps in electronic devices such as transistors.
Microscopic identification of the detrimental defects is crucial
in order to mitigate their impact.  Deep level transient spectroscopy
(DLTS) is a powerful technique for determining the properties of
defects; from an analysis of electrical measurements on a $pn$ junction or Schottky
diode, properties such as the position of the defect level within the band gap, electrical nature (donor or
acceptor), density, and carrier capture cross section of specific defects can be obtained.\cite{peaker2018junction,Henry_Lang_PRB77,Mooney:1998}

Translating this wealth of information to a microscopic identification of a given defect
requires comparison with theoretical or computational models, and first-principles calculations based on density functional theory (DFT)
have proven very helpful.\cite{zhang2016deep, Peaker_FeH_Si_DLTS_APL15, trinh2013negative,reshchikov2017evaluation, coutinho2012electronic}

One of the key quantities measured in DLTS is the activation energy for carrier emission from a defect, $\Delta E_a$.
Defect identification is often based on comparing $\Delta E_a$ with values of the defect ionization energy $\Delta E_i$
determined from zero-temperature first-principles calculations.
However, the underlying theory of DLTS \cite{Henry_Lang_PRB77,Mooney:1998} makes clear that $\Delta E_a$ and $\Delta E_i$ are distinct, and the
use of $\Delta E_i$ can affect the correct identification of a defect.

In the present study we describe a first-principles approach to explicitly determine the activation energies measured in DLTS.
Recent advances have enabled the quantitative prediction of defect levels in the band gap \cite{CGWalle_defects_RMP}
and the ability to accurately describe nonradiative carrier capture.\cite{Alkauskas2014}
We will show that $\Delta E_a$ can significantly differ from $\Delta E_i$ for some defects,
demonstrating the need to explicitly calculate the activation energy in order to correctly identify defects detected by DLTS.
Our analysis is general, but will be illustrated with examples of deep defects in GaN,
a material of high technological relevance because of its applications in solid-state light emitters and power electronics.

\begin{figure}[!t]
\includegraphics[width=8.5cm]{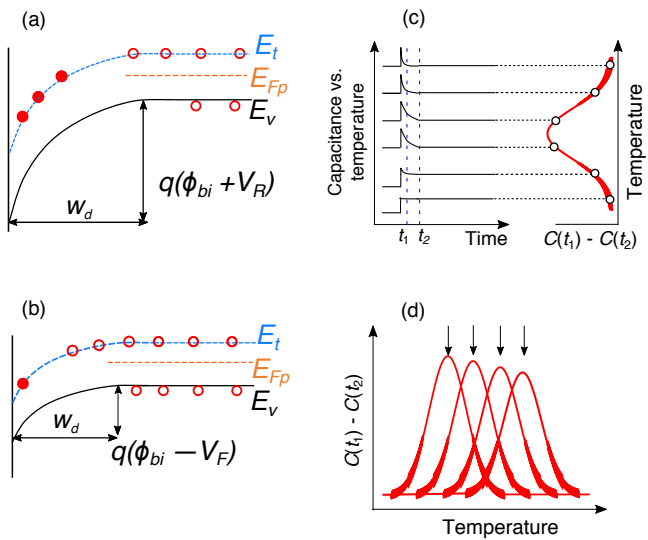}
\caption{Schematic illustration of the DLTS measurement process for a $p$-type Schottky junction with a single
deep acceptor level at an energy $E_{t}$.  $E_{Fp}$ is the quasi-Fermi
level for holes for the junction under bias and $E_{v}$ is the valence-band maximum.
The band diagram is shown under (a) reverse bias, $V_{R}$, and (b) forward bias, $V_{F}$.
$\phi_{bi}$ is the built-in potential at the Schottky junction.
(c) Capacitance measurement as a function of temperature within the DLTS rate window.
(d) Resulting hole emission spectra as a function of temperature obtained for different rate windows.
Panel (c) is adapted from Fig. 6 of Ref.~\onlinecite{DLTS_DVLang_JAP74}.}
\label{fig:dlts}
\end{figure}

Let us consider a defect that acts as a single deep acceptor. A
standard DLTS measurement relies on a $pn$ junction or Schottky diode
that is reverse biased, which establishes a depletion region
that is free of mobile carriers. The band diagram is illustrated
in Fig.~\ref{fig:dlts}(a).\cite{DLTS_DVLang_JAP74} A forward-bias injection pulse is applied, which decreases the width of the
depletion region [Fig.~\ref{fig:dlts}(b)].  Holes from the valence band are captured
nonradiatively into the acceptor level during the injection pulse.
After the pulse is turned off, the depletion width increases to its reverse-bias value and the holes
captured into the acceptor level are re-emitted into the valence band,
which results in a transient change in the capacitance.  Using the
``double boxcar'' technique \cite{DLTS_DVLang_JAP74}, the difference
in the capacitance, $\Delta C$, is measured at two different times
after the injection pulse is turned off [schematically illustrated
in Fig.~\ref{fig:dlts}(c)].  The duration of time within
which the capacitance measurement occurs is termed the emission rate
window.  The change in the capacitance for a given emission rate
window is measured as a function of temperature $T$.
A peak in the capacitance versus $T$ occurs when the rate at which the acceptor level emits a hole, $e_p$, equals the inverse of the emission rate window. Therefore, repeating this measurement with a variety of rate windows, as shown in Fig.~\ref{fig:dlts}(d), yields measurements of $e_p$ versus $T$.

Under equilibrium
conditions, the principle of detailed balance requires that the rate
of hole capture into the acceptor level is equal to the emission rate
of holes into the valence band. Therefore $e_p$ can be written in
terms of the hole capture cross section $\sigma_p$
as\cite{DLTS_DVLang_JAP74}
\begin{equation}
  e_{p}(T) = \frac{\sigma_{p}(T) \langle v_{p}(T) \rangle N_{v}(T)}{g_{v}}\exp\left ( -\frac{\Delta E_{i}(T)}{k_{B}T} \right ) \, ,
\label{eq:balance}
\end{equation}
where $\langle v_{p} \rangle$ is the average thermal velocity of holes
in the valence band, $N_{v}$ is the effective density of states of the
valence band, $g_v$ is the valley degeneracy, and
$\Delta E_i(T)$ is the ionization energy of the defect.  For the case of the deep acceptor,
$\Delta E_i(T)$ is the energy difference between the valence-band
maximum, $E_v$, and the acceptor level.

To obtain the ionization energy and capture cross section from the emission rate versus $T$ data obtained by the procedure in Fig.~\ref{fig:dlts},
 the temperature dependence of the various quantities in Eq.~(\ref{eq:balance}) must be specified.
 The thermal velocity $\langle v_p\rangle$ is given by $\sqrt{\frac{3k_{\text{B}}T}{m_{\text{h}}}}$,
where $m_\text{h}$ is the hole mass.
Assuming parabolic bands, the valence-band density of states is defined as $N_v=2g_v\left(\frac{2\pi m_{\text{h}}^* k_{\text{B}}T}{\hbar^2}\right)^{3/2}$, where $m_\text{h}^*$ is the density-of-states effective mass of holes (1.50 m$_{0}$ for GaN).\cite{santic2003hole, note}
Therefore, $\langle v_p\rangle N_v\propto T^2$.
The standard procedure in DLTS analysis is to assume $\sigma_p$ is independent of temperature and plot ln$(e_p(T)/T^2)$ versus $1/T$,
and then fit to an Arrhenius expression to extract an activation energy $\Delta E_a$.\cite{DLTS_DVLang_JAP74}
This activation energy would coincide with the 0 K ionization energy of the defect if
$\Delta E_i$ and the prefactor of Eq.~(\ref{eq:balance}) were independent of temperature.

However, in reality $\sigma_p$ can have a strong temperature dependence, and therefore a significant contribution to the slope
obtained from a $\ln (e_p(T)/T^2)$ versus 1/T plot \cite{Mooney:1998}.
This contribution must be clarified in order to extract the activation energy from these plots.
In addition, $\Delta E_i$, i.e., the position of the thermodynamic transition level with respect to the band edge,
is also temperature dependent \cite{allen_cardona_tempEg, van1976entropy}.
Both effects are addressed here from first principles.

We perform DFT calculations based on the generalized Kohn-Sham scheme
using the projector-augmented-wave method with the hybrid functional
of Heyd, Scuseria, and Ernzerhof (HSE) \cite{HSE_ref} as implemented in the
VASP code \cite{VASP_ref,VASP_ref2}.  We use a mixing parameter of 0.31
which results in lattice parameters ($a$=3.19 {\AA} and $c$=5.17 {\AA}) and a band gap (3.495 eV) that are close to the
experimental $T$=0 K lattice parameters \cite{maruska1969preparation}
and band gap \cite{freitas2001structural, Nam_Eg_GaN_temp_APL04} of GaN.  The defect calculations are
performed using a 96-atom supercell, a plane-wave basis set with a 400
eV cutoff, and a (2$\times$2$\times$2) Monkhorst-Pack $k$-point grid to sample
the Brillouin zone.  Spin polarization is explicitly included.  All
atomic relaxations are performed consistently with the HSE functional.
Defect formation energies and thermodynamic transition levels were
calculated using the standard formalism \cite{CGWalle_defects_RMP}
with charge-state corrections applied to account for the periodic
supercells \cite{Freysoldt_PSSb}.

Capture coefficients and their temperature dependence are evaluated using the
formalism in Ref.~\onlinecite{Alkauskas2014}.
In order to discuss the temperature dependence of the capture cross
section, we prefer to consider the capture coefficient,
$C_p(T)=\sigma_p(T)\langle v(T)\rangle$, which is a more
general quantity since it can be calculated without assuming a
thermal velocity for the carriers.

To determine the temperature dependence of $\Delta E_i(T)$ we calculate the temperature dependence of the band edges
and the thermodynamic charge-state transition level.
Two mechanisms contribute to the temperature dependence of the band edges.
The contribution due to electron-phonon interactions was evaluated
by using the methodology of Refs.~\onlinecite{Monserrat1, Monserrat2}, on a $4\times4\times4$
$\mathbf{q}$-point grid.
To determine the contribution due to thermal expansion
we used experimental thermal expansion coefficients \cite{maruska1969preparation} to determine the
lattice expansion at a given temperature, and absolute deformation potentials\cite{VdW_defpot_APL97}
to determine the resulting shift in the band edges.
We verified that the calculated cumulative change in the band gap agrees
with experimental measurements, \cite{Nam_Eg_GaN_temp_APL04}
but we emphasize that our procedure allows us to assess the shifts in the individual band edges (valence band versus conduction band).

To determine the temperature dependence of the defect
levels we calculate the zone-center vibrational frequencies of each defect in their different charge states
using HSE with 216-atom supercells using the T=0 K HSE lattice parameters.
The vibrational frequencies were determined for a
set of atoms within 4 \AA~around the defect while the remaining atoms were kept fixed at their
equilibrium positions.
Details of the calculations are provided in the Supplementary Material.
The vibrational frequencies were used to evaluate the vibrational free energy for a
given charge state within the harmonic
approximation [cf. Eq.~(17) in Ref.~\onlinecite{CGWalle_defects_RMP}].
The difference in vibrational free energy between the two charge states was used to determine the
temperature dependence of the transition level.

\begin{figure}[!t]
\includegraphics[width=8.5cm]{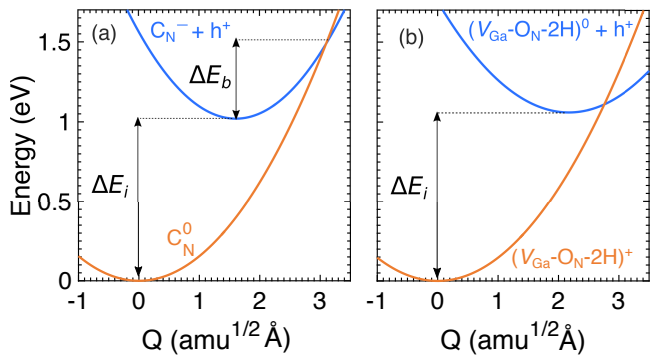}
\caption{
  One-dimensional configuration coordinate diagrams describing hole capture
due to (a) C$_{\rm N}$ and (b) $V_{\rm Ga}$-O$_{\rm N}$-2H in GaN.
In (a) the initial state of the defect is C$_{\rm N}$ in a negative charge
state and the final state is C$_{\rm N}$ in a neutral charge state.
In (b) the initial state is ($V_{\rm Ga}$-O$_{\rm N}$-2H)$^0$
and the final state is ($V_{\rm Ga}$-O$_{\rm N}$-2H)$^+$.
$\Delta E_i$ is energy difference
between the minima of the potential energy surfaces at $T$=0 K and $\Delta E_{b}$
is the classical barrier for the nonradiative capture process.}
\label{fig:ccd}
\end{figure}

We will determine the implications of the $T$ dependence
for two examples of defects in GaN
that have very similar thermodynamic transition levels but different
temperature dependences of their capture coefficient:
carbon on a nitrogen site, $\rm C_{\rm N}$ \cite{lyons_carbon_PRB14},
and a gallium vacancy complex, $V_{\rm Ga}$-O$_{\rm N}$-2H \cite{Lyons_acceptors_PSSb15}.
One-dimensional configuration coordinate diagrams (see Ref.~\onlinecite{Alkauskas2014})
are shown in Fig.~\ref{fig:ccd}.
$\rm C_{\rm N}$ is a deep acceptor, with a (0/$-$) transition level 1.02 eV above $E_v$.
$V_{\text{Ga}}$-O$_{\rm N}$-2H  is a complex based on a gallium vacancy
that exhibits a (+/0) transition level 1.06 eV above $E_v$ \cite{Lyons_acceptors_PSSb15}.

For purposes of determining the capture coefficient, the
initial state of the system consists of a hole in the valence band and a
negatively charged acceptor; see the potential energy surface labeled
$\rm C_{\rm N}^{-}+h^{+}$ in Fig.~\ref{fig:ccd}(a).  Capture occurs when
the system traverses to the potential energy surface corresponding to
the neutral acceptor, $\rm C_{\rm N}^0$.
The difference in energy of the minima of
the two curves is the thermodynamic charge-state transition level referenced to $E_v$,
and corresponds to the ionization
energy $\Delta E_i$ from Eq.~(\ref{eq:balance}). At high
temperatures the capture process
occurs by surmounting the ``classical'' barrier, \cite{Henry_Lang_PRB77} $\Delta E_b$,  obtained
from the intersection point of the curves in the configuration coordinate diagram;
at low temperatures, the transition rate is dominated by
quantum-mechanical tunneling
\cite{Alkauskas2014}.
The classical barrier $\Delta E_{b}$ is 490 meV for $\rm C_{\rm N}$ and 49 meV for $V_{\rm Ga}$-O$_{\rm N}$-2H.

\begin{figure}[!htb]
\includegraphics[width=8.5cm]{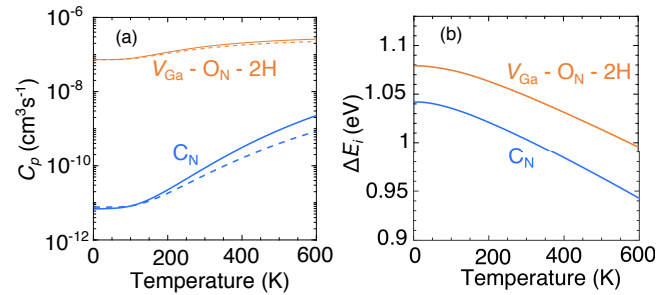}

\caption{
(a) Hole capture coefficient versus temperature for $\rm C_{\rm N}$ and
$V_{\rm Ga}$-O$_{\rm N}$-2H in GaN.  The dashed lines
are based on constant $T$=0 K values for $\Delta E_i$.
The solid lines take the temperature dependence of $\Delta E_i$, as shown in (b),
into account.  (b) Variation in the ionization energy of
$\rm C_{\rm N}$ and $V_{\rm Ga}$-O$_{\rm N}$-2H as a function of temperature, as described in the text.  }
\label{fig:cp}
\end{figure}

The large difference in these classical barriers is reflected in our
results for the temperature dependence of the hole capture coefficients
in Fig.~\ref{fig:cp}(a) (dashed lines).
We focus on the temperature range up to 600 K;
DLTS measurements on GaN are limited to this temperature to prevent degrading of the metal contacts.
The hole capture coefficient of $\rm C_{\rm N}$
changes by two orders of magnitude as the temperature increases from 0 K to 600 K,
while for $V_{\rm Ga}$-O$_{\rm N}$-2H the temperature dependence is much more modest.
The results shown in dashed lines in Fig.~\ref{fig:cp}(a) assume that the $\Delta E_i$
are fixed to their $T$=0 values.
These $\Delta E_i$ values correspond to charge-state transition levels obtained from static-lattice calculations
of a zero-temperature DFT calculation.
Both dashed curves have an Arrhenius form at high $T$, and
have a weak temperature dependence as $T\rightarrow 0$. \cite{Alkauskas2014}
In reality, the distance between $E_v$ and the defect level shrinks as $T$ increases, and so $\Delta E_i(T)$
is reduced, as shown in Fig.~\ref{fig:cp}(b).
Inclusion of this additional effect enhances the dependence of $C_p$ on $T$,
as shown by the solid curves in Fig.~\ref{fig:cp}(a).

It is commonly assumed\cite{Mooney:1998} that the capture coefficient has a temperature dependence
given by $C_p = C_{\infty} \exp(-\Delta E_b / k_{B}T)$.
Our results in Fig.~\ref{fig:cp}(a) show that the description in terms of a temperature-independent classical barrier
is too simple to capture the actual temperature dependence of $C_p $.
At low temperatures, the Arrhenius form would imply that $C_p$ goes to zero as
$T\rightarrow 0$ K, but in reality $C_p$ remains finite because of quantum-mechanical tunneling.
At high $T$, the behavior is also non-exponential, caused by the temperature dependence of $\Delta E_i$.
Hence, in a quantum-mechanical treatment of nonradiative capture of carriers by defects one
should consider an effective barrier to describe the temperature dependence of such processes \cite{Alkauskas2014}.
Unlike the classical capture barrier $\Delta E_b$, the effective barrier $\Delta E_b^{\prime}(T)$ is temperature dependent
resulting in $C_p$ deviating from purely exponential behavior.

\begin{figure}[!tb]
\includegraphics[width=7.5cm]{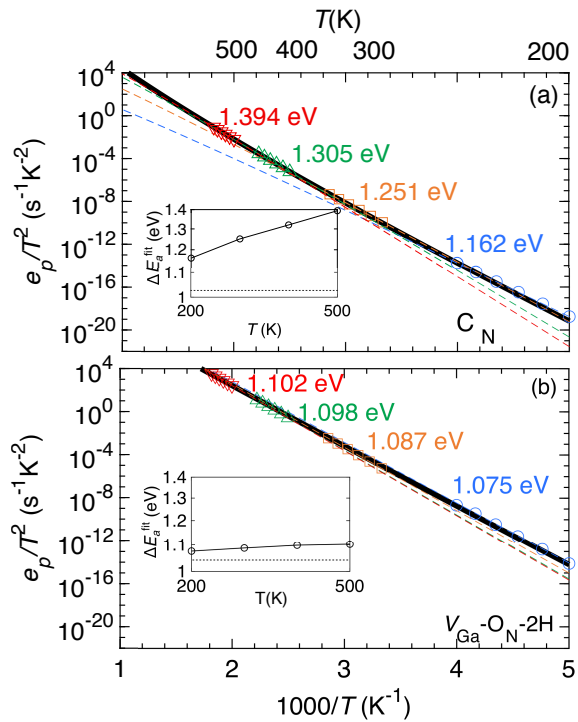}
\caption{Calculated hole emission rates for (a) $\rm C_{\rm N}$ and (b)
$V_{\rm Ga}$-O$_{\rm N}$-2H in GaN (solid lines).  The dashed lines are least-squares fits to
Eq.~(\ref{eq:fit}), with the thick band of symbols indicating the
temperature ranges over which the fit was performed: 200--250 K, 300--350 K, 400--450 K and
500--550 K. Extracted activation energies $\Delta E_a$ are shown alongside each fit and plotted
as a function of temperature in the inset.  The zero-temperature ionization energy
for each defect is illustrated with a horizontal dashed line.}
\label{fig:ep_fit}
\end{figure}

We now use our values of $C_p$
[Fig.~\ref{fig:cp}(a)]
to calculate the emission rate based on Eq.~(\ref{eq:balance}).
We mentioned before that the common practice in DLTS analysis is to plot $\ln(e_p/T^2)$ versus $1/T$,
based on a lack of information about the temperature dependence of $\sigma_{p}$ and the fact that $\langle v_p\rangle N_v\propto T^2$.
Therefore in Fig.~\ref{fig:ep_fit} we plot $\ln(e_p/T^2)$.

When plotted over this large temperature range, the calculated emission rates (black solid lines in Fig.~\ref{fig:ep_fit}) clearly
deviate from linearity, reflecting the non-Arrhenius behavior of $C_p$ as well as the temperature dependence of $\Delta E_i$.
It is important to note that even if $\ln(e_p/T^{2})$ would be linear, the slope still
does not correspond to the ionization energy, as it includes the capture barrier.

DLTS measurements are carried out over a limited temperature range, typically about 50 K, and the data are then fitted to an Arrhenius expression.
Based on the data as plotted in our Fig.~\ref{fig:ep_fit}, we fit to an expression:
\begin{equation}
e_p/T^{2} = e_0^{\text{fit}}{\rm \exp} (-\Delta E_{a}^{\text{fit}}/k_{B}T) \, .
\label{eq:fit}
\end{equation}
We can thus determine the activation energy $\Delta E_{a}^{\text{fit}}$ that would be extracted from a typical DLTS measurement by fitting 
over a finite temperature range similar to the one probed in experiments (dashed curves in  Fig.~\ref{fig:ep_fit}).
We find that for C$_{\rm N}$, the fitted activation energies increase from
1.162 eV to 1.394 eV, depending on the temperature range for which the fit is performed [Fig.~\ref{fig:ep_fit}(a)]. For $V_{\rm Ga}$-O$_{\rm N}$-2H,
the explicit calculations are much closer to a simple Arrhenius behavior, and hence there is little variation in
the activation energies extracted over different temperature ranges [Fig.~\ref{fig:ep_fit}(b)].

The activation energy is temperature dependent and the deviation
between the ionization energy and activation energy is pronounced at higher temperatures.
Our calculations highlight that the difference between the activation energy and the ionization energy can be large: up to 0.4 eV for C$_{\text{N}}$.
The activation energy obtained from an Arrhenius analysis of the emission rate differs from the 0 K ionization energy of the defect for two reasons:
first, because the activation energy also includes a capture barrier, and second, because the ionization energy itself is temperature dependent, due to the temperature dependence of the band edges and of the defect transition level.
Activation energies extracted from DLTS should therefore not be simply interpreted as ionization energies, and simple comparisons with
first-principles ionization energies could lead to incorrect identification of defects.

As mentioned above, the typical procedure (which we have followed in this paper) is to plot the results of DLTS experiments as $\ln(e_p/T^2)$,
and perform an Arrhenius analysis assuming that $\sigma_p$ has a temperature-independent prefactor [see Eq.~\ref{eq:balance}].
It has been shown, however, that at high temperature the preexponential factor in $\sigma_p$ has a $1/T$
dependence [{\it cf.} Eq.~(28) in Ref.~\onlinecite{Henry_Lang_PRB77} and Eq.~(61a) in Ref.~\onlinecite{ridley1978multiphonon}].
At high temperature one should therefore perform an Arrhenius analysis of  $\ln(e_p/T)$, and this would be important to recover
the value of the classical barrier $\Delta E_b$.
Indeed, we find that including this $1/T$ dependence when fitting our first-principles calculations of capture coefficients
at high temperatures ($T >$ 1200 K) results in $\Delta E^{\prime}_{b} \rightarrow \Delta E{_b}$ (the classical barrier at fixed $\Delta E_i$, see Fig.~2), as expected.
However, the typical temperature range over which DLTS experiments are performed
does not reach this “high-temperature” limit, and therefore $\ln(e_p/T^2)$ is an acceptable approximation.

In summary, we have shown DLTS activation energies are temperature-dependent and should not be compared directly with first-principles calculations of ionization energies of defects.  
Using first-principles calculations we determined the temperature dependence of nonradiative carrier capture and the ionization energy of defects in GaN and demonstrated 
how they yield activation energies that can differ greatly from the 0 K ionization energy of the defect.  
The C$_{\text{N}}$ and $V_{\rm Ga}$-O$_{\rm N}$-2H defects we considered
in this study are examples of positive-$U$ defects where we determined the activation energy due to thermal emission from a single thermodynamic transition level.   
Our conclusions on the temperature dependence of activation energy will also apply in the case of more complex situations 
such as defects with two thermodynamic transition levels that are amenable to ionization in DLTS. 
In the case of a positive-$U$ center, thermal emission due to both thermodynamic transition levels would be observed in a DLTS measurement.  
Our formalism can be applied to determine the activation energy of both transitions separately. 
In the case of a negative-$U$ center, the DLTS transient is determined by the slower of the two carrier emission process. 
Thus, one would observe a single peak with an activation energy that corresponds to the slower emission process, to which our analysis is equally applicable.  
Hence, our analysis of these quantities is general and can be applied to accurately determine defect activation energies when comparing to DLTS measurements.

\textbf{Supplementary Material:} See Supplementary Material for the details of calculations of the vibrational properties of defects
and a comparison between the calculated and experimental temperature dependence of the band gap of GaN.

\textbf{Acknowledgements: }
D.\ W.\ was supported by the National Science Foundation (NSF) under Grant No. DMR-1434854.
J.\ S.\ was supported by the U.\ S. Department of Energy, Office of Science, Basic Energy Sciences, under Award No. DE-SC0010689.
B.\ M.\ acknowledges support from the Winton Programme for the Physics of
Sustainability, and from Robinson College, Cambridge, and the Cambridge
Philosophical Society for a Henslow Research Fellowship.
A.\ A.\ was supported by Marie Sk{\l}odowska-Curie Action of the European Union (project N\textsc{itride}-SRH, grant No.~657054).
The Flatiron Institute is a division of the Simons Foundation.
Computational resources were provided by the Extreme Science and Engineering Discovery Environment (XSEDE), support by NSF (ACI-1053575).

%

\begin{thebibliography}{28}
\bibitem [{not()}]{note}%
  \BibitemOpen
  \href@noop {} {\bibinfo  {journal} {{ Use of this $m_\text{h}^*$ value in
  $N_v$ does not take splitting between the valence bands into account. We have
  verified that this approximation does not impact our conclusions.}}\
  }\BibitemShut {NoStop}%
\end{thebibliography}
\clearpage
\pagebreak
\widetext
\setcounter{equation}{0}
\setcounter{figure}{0}
\setcounter{table}{0}
\renewcommand{\thefigure}{S\arabic{figure}}
\renewcommand{\theequation}{S\arabic{equation}}
\renewcommand{\thesection}{S\arabic{section}}

\begin{center}
\textbf{\large Supplementary Material: ``Defect identification based on first-principles calculations for deep level transient spectroscopy''}
\end{center}
\begin{center}
\section*{Vibrational and thermodynamic properties of defects}
\end{center}

\subsection{Procedure for calculating the vibrational contribution to the defect level}

We are interested in determining the temperature dependence of the defect ionization energy, $\Delta E_i$.
One contribution arises from the temperature dependence of the valence-band edge due to electron-phonon interactions and thermal expansion;
the calculation of this contribution is described in the main text.
A second contribution to the temperature dependence is due to vibrational entropy of the defect, which shifts the thermodynamic transition level.
We can evaluate this contribution by calculating the vibrational free energy of the defect in each of the relevant charge states; the difference yields the shift in the transition level.

Calculations of free energy require the evaluation of vibrational frequencies.
The zone-center vibrational properties of the defect are calculated in a defect supercell, where one defect is embedded in a large
volume of host material and is periodically repeated.
A finite-difference scheme is used to obtain vibrational frequencies.
The $T = 0$ equilibrium HSE lattice parameters of GaN were used.
In the course of our convergence tests, we found that an energy convergence criterion of 10$^{-7}$ eV needs to be applied for these calculations; this is a much more stringent criterion than the default value of 10$^{-4}$ eV.

The vibrational frequencies are used to determine the vibrational free energy $F_{ph}$ for the defect in a charge state $q$ within the harmonic approximation \cite{SM_CGWalle_defects_RMP}:
\begin{equation}
F_{ph}=\sum_{i}\left[\frac{1}{2}\hbar\omega_i + k_B T \ln \left \{ 1 - \exp \left(-\frac{\hbar\omega_i}{k_BT} \right) \right \} \right] \, .
\label{free}
\end{equation}
The impact of vibrations on the
temperature dependence of the thermodynamic level is determined by the difference in vibrational
free energy, $\Delta F_{ph}$, between the two charge states of the thermodynamic level.  For example, for C$_{\rm N}$ this is the
difference between $F_{ph}$ of the negative and neutral charge states, as defined in Eq.~(\ref{eq:del_fph}):
\begin{equation}
\Delta F_{ph} = F_{ph}(C_{\rm N}^{q=1-}) - F_{ph}(C_{\rm N}^{q=0}) \, .
\label{eq:del_fph}
\end{equation}
For $V_{\text{Ga}}$-O$_{\rm N}$-2H $\Delta F_{ph}$ is the difference between F$_{ph}$ in the neutral and positive charge state.

In principle, we would like to include the vibrational properties of all of the atoms in the supercell in the calculation of $F_{ph}$.
However, for the supercell sizes that are needed (see our convergence tests below), this is computationally intractable when using a hybrid functional such as HSE \cite{SM_HSE_ref}
using the convergence criteria that we have identified to be necessary to obtain $\Delta F_{ph}$ with acceptable accuracy.
One approach would be to use a less demanding functional such as the generalized gradient approximation of Perdew, Burke, and Ernzerhof (PBE) \cite{SM_PBE}. This is the approach we have used for conducting the benchmark and supercell-size convergence tests reported below.

There is a concern, however, that PBE may not capture all of the relevant properties.  Indeed, certain defect configurations lead to charge localization that is properly described only when using a hybrid functional~\cite{SM_CGWalle_defects_RMP}, and the atomic relaxations that accompany this localization may induce significant changes in vibrational properties.
We have therefore developed an alternative procedure for calculating $\Delta F_{ph}$, in which only atoms in the vicinity of the defect center are included in the calculation of the vibrational properties.  Based on the tests reported below, we have found that including atoms within 4 \AA~of the defect center (corresponding to first and second nearest neighbors) is sufficient to calculate $\Delta F_{ph}$ to an acceptable degree of accuracy.
The remaining atoms outside of this 4-{\AA} radius are kept fixed at their relaxed equilibrium positions.
In practice, we have used 216-atom supercells where the atomic coordinates in each charge state were relaxed using HSE.  For each charge state $q$ of the defect, the vibrational frequencies of the atoms within this 4-{\AA} radius are calculated with the HSE hybrid functional using finite differences.
$\Delta F_{ph}$ is then obtained based on Eq.~(\ref{eq:del_fph}).

Our convergence tests are detailed in the subsections below.

\subsection{Benchmark calculations for vibrational properties}
\label{sec:supercell}

Since full HSE calculations of vibrational properties in a sufficiently large supercell are intractable, we have developed an alternative procedure
based on calculating vibrational properties for a subset of atoms.
In order to check the accuracy of the procedure, we need to have a benchmark value for $\Delta F_{ph}$, calculated by including vibrations for all the atoms in the supercell.
Since such calculations are not feasible with HSE, we instead used PBE \cite{SM_PBE}
to determine the convergence of vibrational frequencies and free energies as a function of supercell size.
We assume that the convergence properties as a function of supercell size will be similar for HSE.

\begin{figure}[!h]
\includegraphics[width=12.0cm]{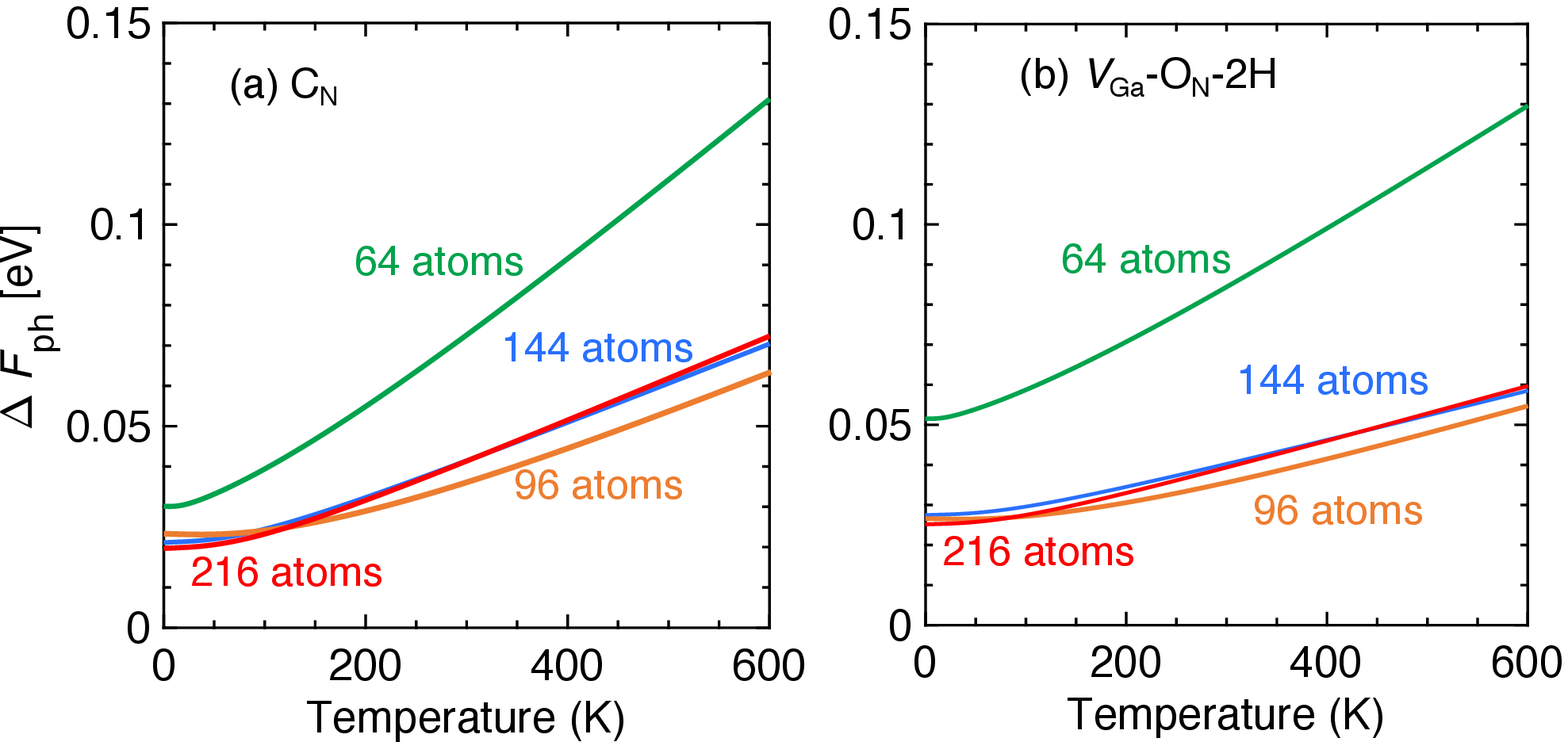}
\caption{$\Delta F_{ph}$ for (a) C$_{\rm N}$ and (b) $V_{\text{Ga}}$-O$_{\rm N}$-2H
as a function of temperature for 64, 96, 144 and 216 atom wurtzite cells including all vibrational modes determined with the PBE functional.}
\label{fig:deltafph-pbe}
\end{figure}

Figure~\ref{fig:deltafph-pbe} shows the convergence of $\Delta F_{ph}$ for the defects in GaN as a function of supercell size.
This allows us to determine the supercell size where $\Delta F_{ph}$ for an isolated defect is not impacted by finite-size effects.
We consider supercell sizes that range from 64 atoms to 216 atoms.
For each of these calculations PBE lattice parameters of GaN were used and all atomic relaxations were performed within PBE.
$\Delta F_{ph}$ as obtained using PBE vibrational modes is shown in Fig.~\ref{fig:deltafph-pbe}.
We conclude that supercells larger than 144 atoms are needed to obtain converged results.

\subsection{Vibrational properties of defects using a finite number of atoms}

\begin{figure}[!h]
\includegraphics[width=12.0cm]{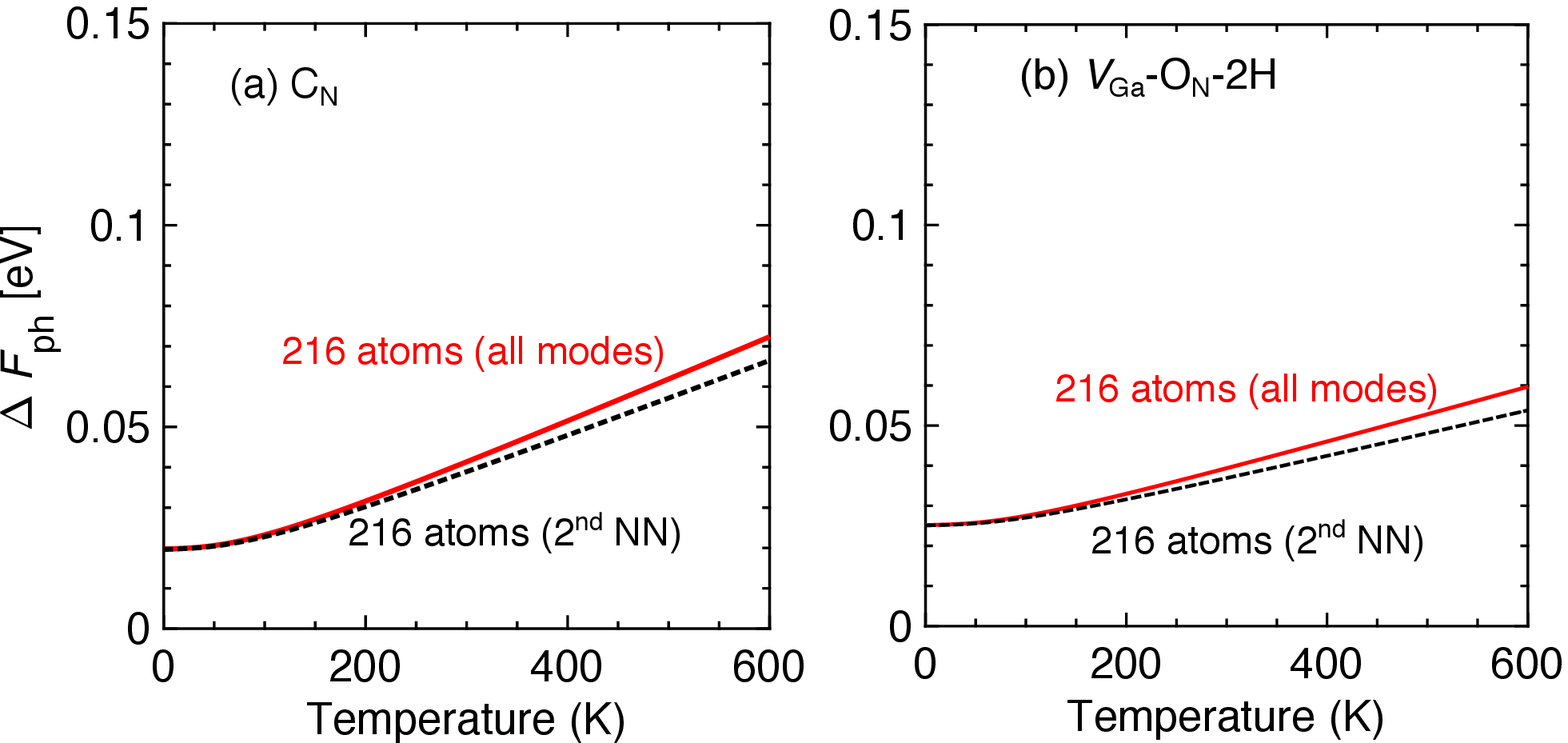}
\caption{$\Delta F_{ph}$ including all vibrational modes versus vibrational modes contributed by atoms lying within 4~{\AA} of the defect center
[up to 2$^{\rm nd}$ nearest neighbors (NN)] for (a) C$_{\rm N}$ and (b) $V_{\text{Ga}}$-O$_{\rm N}$-2H determined with the PBE functional.}
\label{fig:shell-delph}
\end{figure}

We have developed a procedure to obtain $\Delta F_{ph}$
based on the vibrational properties of a subset of atoms within 4~{\AA} around the defect site.
The set of atoms lying within 4 {\AA} of the defect site corresponds to including atoms that constitute the defect center, plus their first and second nearest neighbors.
The remaining atoms in the supercell are kept fixed at their equilibrium positions.
Vibrations of atoms lying outside this 4 \AA~sphere will definitely contribute to the vibrational free energy; however, our hypothesis is that these contributions will be very similar for different charge states $q$, and hence will cancel in $\Delta F_{ph}$ [see Eq.~\ref{eq:del_fph}].
We have verified this approach by calculating $\Delta F_{ph}$ in a 216-atom supercell for C$_{\rm N}$ and $V_{\text{Ga}}$-O$_{\rm N}$-2H, and comparing
the result obtained based on the vibrational modes of atoms lying within 4 {\AA} of the defect site with the result obtained based on the vibrational modes of all atoms in the supercell.
The PBE functional was used for these tests, and the results are illustrated in Fig.~\ref{fig:shell-delph}.

From Fig.~\ref{fig:shell-delph} it is evident that a majority of the change in the vibrational free energy is
captured by the vibrational modes of the atoms within this small volume.  This now makes it feasible to determine $\Delta F_{ph}$ for these defects using the HSE hybrid functional.

\begin{figure}[!h]
\includegraphics[width=16.5cm]{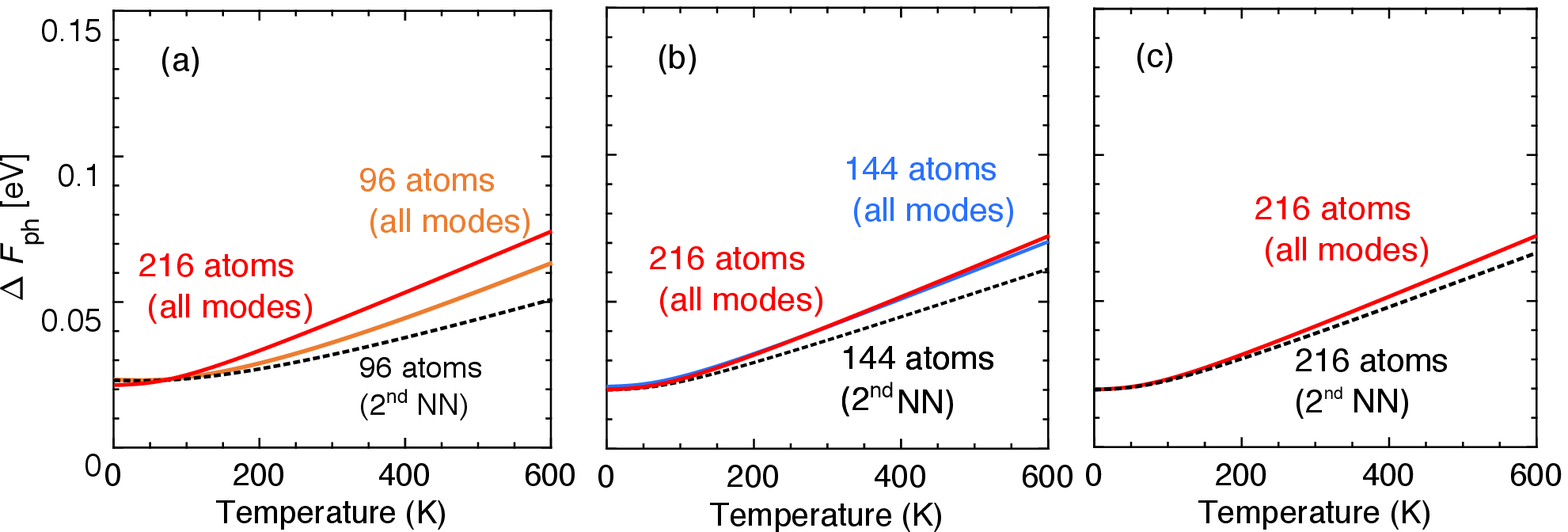}
\caption{Comparison of $\Delta F_{ph}$ for C$_{\rm N}$ as determined with the PBE functional, for different supercell sizes. (a) 96 atoms; (b) 144 atoms; (c) 216 atoms.
In each case the result obtained vibrational modes contributed by atoms lying within 4~{\AA} of the defect center (up to 2$^{\rm nd}$ NN) is compared with the result including vibrational modes for all atoms within the same-size supercell, as well as with the result including vibrational modes for all atoms within a 216-atom supercell (which serves as our benchmark).}
\label{fig:4Aconv}
\end{figure}

One question still to be addressed relates to the supercell size that is needed for this procedure to yield acceptable results.  One may think that, since atoms outside a 4-{\AA} radius are kept fixed, a relatively small supercell might suffice.  We have carried out tests for 96-, 144-, and 216-atom supercells, as shown in Fig.~\ref{fig:4Aconv}.  The comparison shows that 216-atom supercells yield the best results.  While atoms outside the 4-{\AA} radius are kept fixed in the process of calculating vibrational properties, these vibrational properties turn out to be quite sensitive to the supercell size.  This may be because of a sensitivity to the details of the atomic relaxations (which may not be fully converged in a 96-atom cell).
More likely it is due to interactions between neighboring supercells in the course of the evaluation of vibrational frequencies.  Such a spurious interaction may be present particularly in the case of nonzero charge states, especially since the suppression of relaxation of atoms outside the 4-{\AA} radius also suppresses screening.  These spurious effects diminish with increasing supercell size.

Based on these tests, 216-atom supercells were used with the HSE lattice parameters of GaN, including complete relaxation of all atoms with HSE, and the
vibrational properties were also determined with HSE for atoms within 4 {\AA} of the defect site.  The results are shown in Sec.~\ref{sec:HSEresults}.

\subsection{Impact of thermal expansion on vibrational properties}

\begin{figure}[!h]
\includegraphics[width=12.0cm]{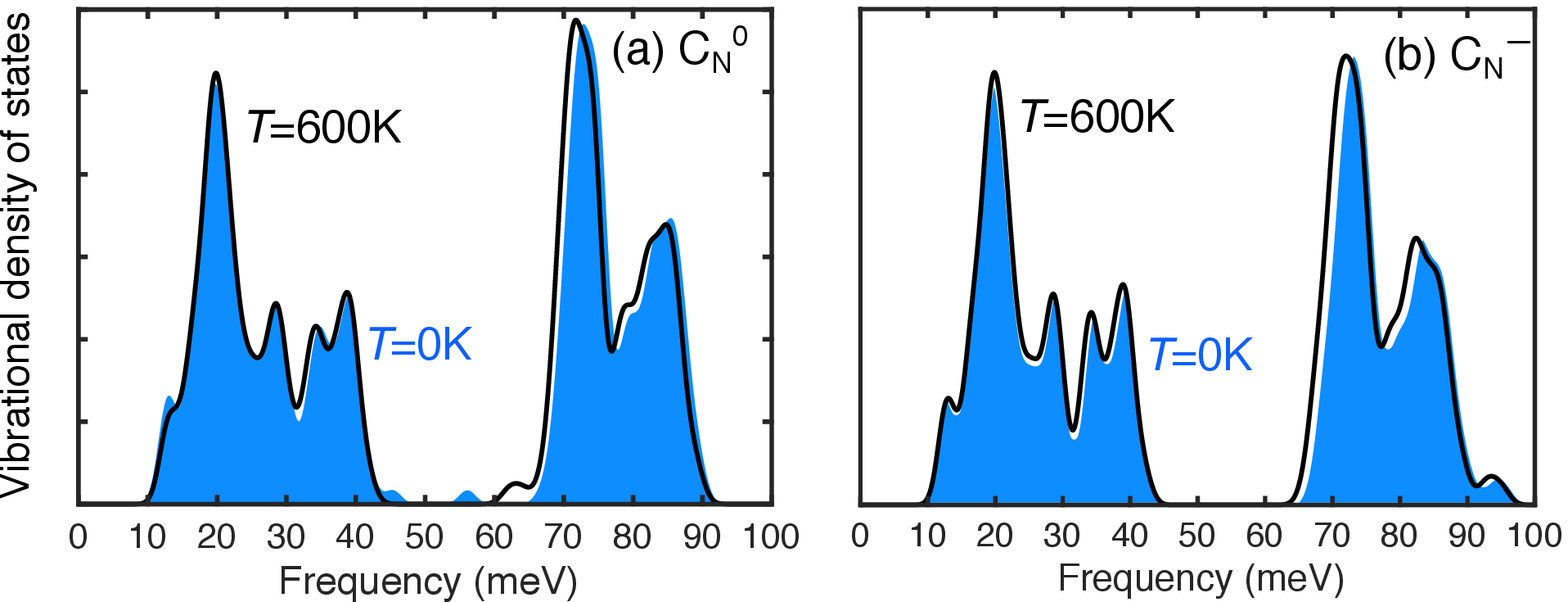}
\caption{Vibrational density of states for the (a) neutral and (b) negative charge state of C$_{\rm N}$, calculated  using the equilibrium ($T$=0 K) (shaded blue)
and expanded lattice parameters (corresponding to thermal expansion at $T$=600 K) (black line).  Vibrational frequencies are calculated within PBE.  }
\label{fig:pbe-thermalexpansion}
\end{figure}

Finite temperature leads to thermal expansion of the GaN lattice; in turn this can impact the vibrational frequencies of the
defect.  In this study we have used the equilibrium 0 K HSE lattice parameters to determine the
vibrational frequencies.  To justify this approximation we compare the vibrational frequencies of C$_{\rm N}$ and $V_{\text{Ga}}$-O$_{\rm N}$-2H
obtained with a 216-atom GaN supercell with 0 K PBE lattice parameters
with a 216 atom supercell with expanded lattice parameters (based on the thermal expansion coefficients of GaN \cite{SM_maruska1969preparation} and a temperature of 600 K).
We find thermal expansion to have a minor impact on the vibrational frequencies of the defects.
As an example, the
vibrational density of states for C$_{\rm N}$ in the neutral and negative charge states
between the 0 K and 600 K lattice constants is illustrated in Fig.~\ref{fig:pbe-thermalexpansion}.

The difference in vibrational properties has a very small impact on $\Delta F_{ph}$.
$\Delta F_{ph}$ calculated with the 600 K lattice parameters differs by only 4 meV from the value calculated with $T=0$ lattice parameters for C$_{\rm N}$, and by 3 meV for $V_{\text{Ga}}$-O$_{\rm N}$-2H.
These differences in $\Delta F_{ph}$ are significantly smaller than the other temperature-dependent quantitites we consider
in this study.
We conclude that we can neglect lattice expansion in the calculation of the vibrational free energy, and we have calculated the vibrational frequencies for both defects using
the equilibrium 0 K lattice parameters.

\subsection{ Vibrational free-energy contribution to the thermodynamic transition level: HSE results}
\label{sec:HSEresults}

\begin{figure}[!h]
\includegraphics[width=12.0cm]{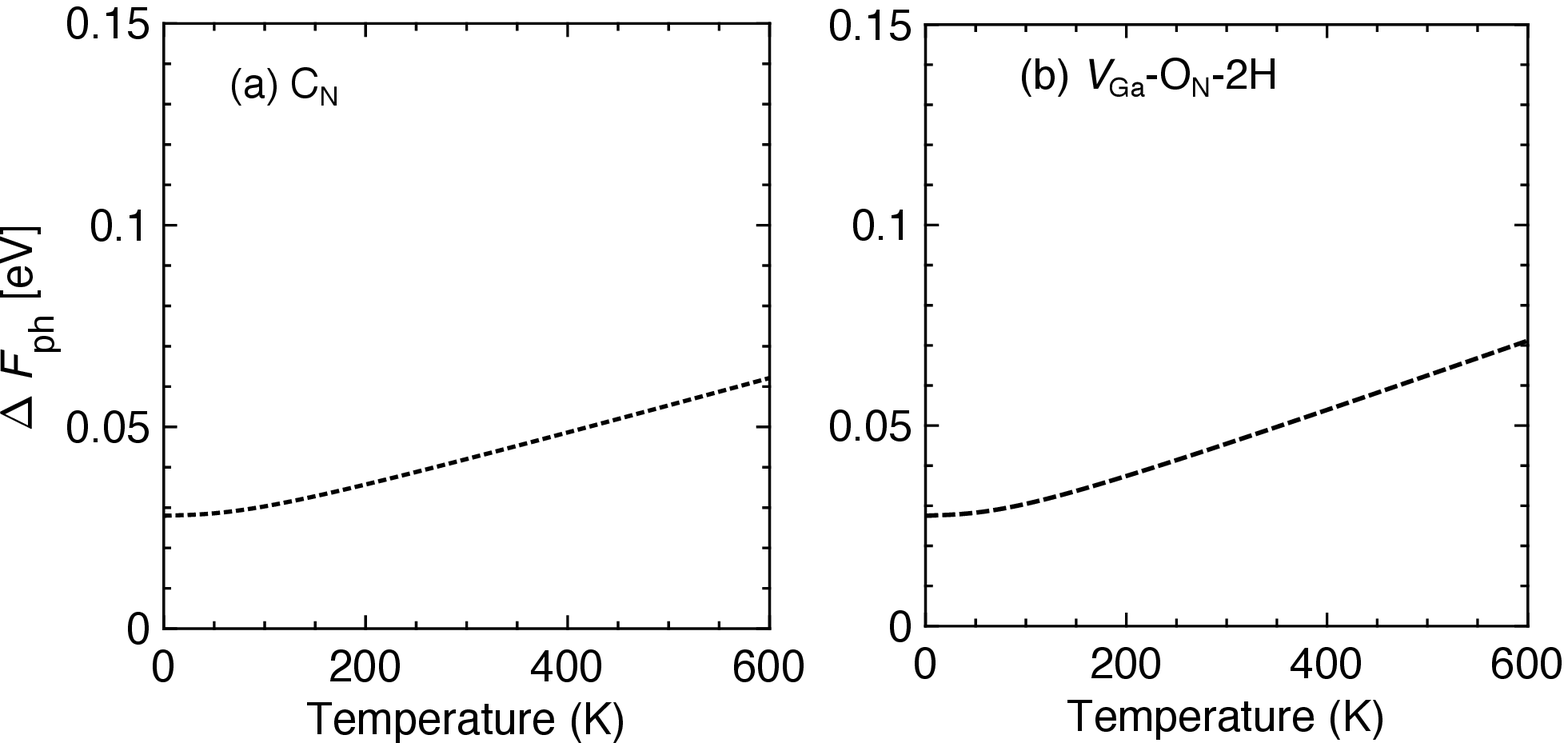}
\caption{$\Delta F_{ph}$ for (a) C$_{\rm N}$ and (b) $V_{\text{Ga}}$-O$_{\rm N}$-2H determined with the HSE
function for a atoms lying within 4~{\AA} of the defect center in a 216-atom supercell.}
\label{fig:shell-delph-hse}
\end{figure}

Figure~\ref{fig:shell-delph-hse} shows $\Delta F_{ph}$ for C$_{\rm N}$ and
$V_{\text{Ga}}$-O$_{\rm N}$-2H in GaN as a function of temperature, obtained with HSE calculations within 216 atom supercells as described above:
the T=0 K HSE lattice parameters were used and only atoms within 4 \AA~of the defect center were used to compute the vibrational
frequencies.  The remaining atoms were kept fixed at their equilibrium positions.

These results were used to determine the impact of vibrational free energy on the temperature dependence of the defect transition level.
The results in Fig.~\ref{fig:shell-delph-hse}, combined with our calculated temperature dependence of the band edges, were used to
determine the temperature dependence of the ionization energy $\Delta E_i$ [cf. Fig.~3(b) in the main text].
\section*{Temperature dependence of the G\MakeLowercase{a}N band gap}
\begin{figure}[!h]
\includegraphics[width=8.0cm]{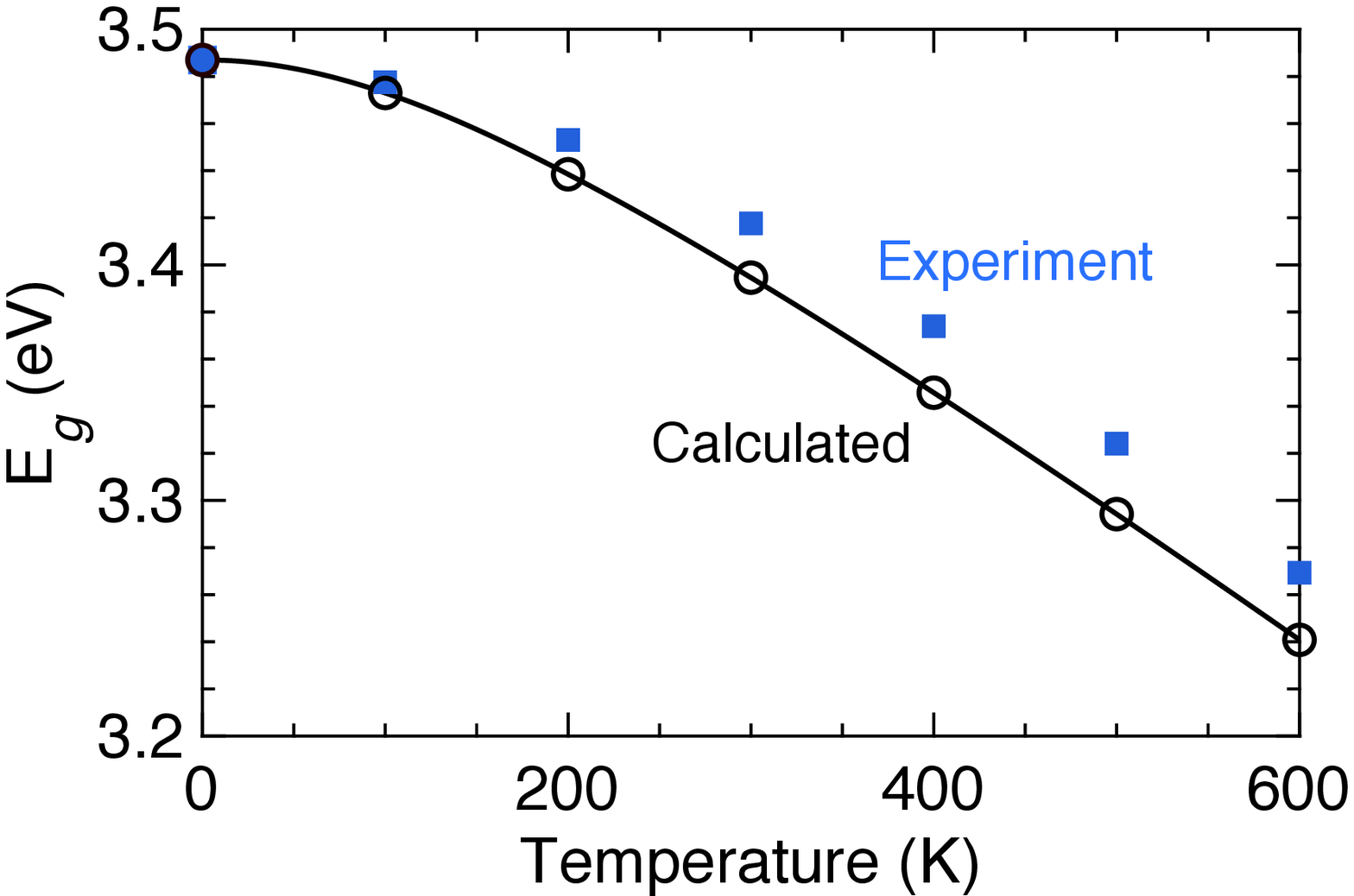}
\caption{Calculated (black) and experimentally measured \cite{SM_Nam_Eg_GaN_temp_APL04} (blue squares) change in the band gap of GaN as a function of temperature.}
\label{fig:egtemp}
\end{figure}

We are interested in the temperature dependence of the band gap and band edges of GaN since these changes in the
electronic structure impact the capture coefficients and emission rates.
Our calculations take into account the role of electron-phonon interactions and thermal expansion. 
In Fig.~\ref{fig:egtemp} we show the change in the band gap as a function of temperature, 
calculated using the methodology described in the main text.
We compare these results to the experimentally measured change in the band gap of GaN as reported in Ref.~\onlinecite{SM_Nam_Eg_GaN_temp_APL04}.

We note that the GaN layers in the experimental measurements of Ref.~\onlinecite{SM_Nam_Eg_GaN_temp_APL04} were grown on sapphire and thus experience a certain amount of strain, which will vary with temperature.  
Such strain effects are not included in our calculations and may be responsible for the difference between experimental and calculated results.

\end{document}